\documentclass[12pt]{article} %[reprint,aps]{revtex4-1}
\usepackage{epsfig, amsmath, amssymb}
\usepackage{graphicx,psfrag} 
\newcommand{\be}{\begin{equation}}
\newcommand{\ee}{\end{equation}}
\newcommand{\ben}{\begin{equation}}
\newcommand{\een}{\end{equation}}
\newcommand{\bea}{\begin{eqnarray}}
\newcommand{\eea}{\end{eqnarray}}
\newcommand{\bA}{\begin{array}}
\newcommand{\eA}{\end{array}}
\newcommand{\bc}{\begin{center}}
\newcommand{\ec}{\end{center}}
\newcommand{\al}{\alpha}

\newcommand{\ra}{\rightarrow}
\newcommand{\del}{\partial}

\newcommand{\ie}{{\it i.e.}}
\newcommand{\eg}{{\it e.g.}}

\newcommand{\vx}{{\vec x}}

\newcommand{\cO}{{\cal O}}
\newcommand{\cI}{{\cal I}}

\begin{document}

%\ifeprint
%\fi

\begin{titlepage}
%\vspace{30mm}

\bc

%\hfill  {TIFR/TH/09-12} \\
\hfill % {\tt arXiv:0909.4731 [hep-th]} 
\\         [22mm]
%%X\vfill

{\Huge $dS/CFT$ at uniform energy density \\ [2mm] 
and a de Sitter ``bluewall''} 
\vspace{16mm}

{\large Diptarka Das$^a$, Sumit R. Das$^a$ and K.~Narayan$^b$} \\
\vspace{3mm}
{$^a$\small \it Department of Physics and Astronomy, \\}
{\small \it University of Kentucky, Lexington, KY 40506, USA.\\}
\vspace{3mm}
{$^b$\small \it Chennai Mathematical Institute, \\}
{\small \it SIPCOT IT Park, Siruseri 603103, India.\\}
%{\small Email: \ narayan@cmi.ac.in}\\

\ec
\medskip
\vspace{40mm}

\begin{abstract}
We describe a class of spacetimes that are asymptotically de Sitter in
the Poincare slicing. Assuming that a $dS/CFT$ correspondence exists,
we argue that these are gravity duals to a CFT on a circle leading to uniform
energy-momentum density, and are equivalent to an analytic
continuation of the Euclidean $AdS$ black brane.  These are solutions
with a complex parameter which then gives a real energy-momentum
density.  We also discuss a related solution with the parameter 
continued to a real number, which we refer to as a de Sitter
``bluewall''. This spacetime has two asymptotic de Sitter universes
and Cauchy horizons cloaking timelike singularities. We argue that the
Cauchy horizons give rise to a blue-shift instability.
\end{abstract}

\end{titlepage}

%\newpage 
%{\tiny %footnotesize
%\begin{tableofcontents}
%\end{tableofcontents}
%}

\vspace{5mm}

\section{Introduction and summary}

One  version of the  $dS/CFT$ correspondence 
\cite{Strominger:2001pn,Witten:2001kn,Maldacena:2002vr} states that 
quantum gravity in de Sitter 
space is dual to a Euclidean CFT living on the boundary $\cI^+$ or $\cI^-$. 
More specifically, the partition function of the CFT with specified sources $\phi_{i0}(\vx)$ coupled to operators $\cO_i$ is identified with the wavefunctional of the bulk theory as a functional of the boundary values of the fields dual to $\cO_i$ given by $\phi_{i0}(\vx)$. In the semiclassical regime this becomes
\ben
\Psi[\phi_{i0}(\vx)] = {\rm exp} \left[ iI_{cl}(\phi_{i0}) \right]
\label{one}
\een
where we need to impose regularity conditions on the cosmological horizon. 
This has been developed further in \cite{Harlow:2011ke}.
Unlike AdS/CFT, there are few concrete realizations of dS/CFT (for a
recent proposal see \cite{Anninos:2011ui} and \eg\ \cite{Ng:2012xp,
Anninos:2012qw,Das:2012dt,Anninos:2012ft,Anninos:2013rza,
Das:2013qea,Banerjee:2013mca,Chang:2013afa} for related work). 
Nevertheless, it is interesting to explore the consequences of such a 
correspondence, assuming it exists. 

In this note we address the question of what the bulk dual is
of a euclidean CFT with  {\em constant} spatially uniform energy-momentum
density. One way to achieve this is to put the CFT on a circle.
It is well known that the dual CFT
to de Sitter space cannot be a usual unitary (more precisely
reflection-positive) quantum field theory. The bulk dual of such a
theory would be euclidean AdS. Such a CFT on a circle has a uniform 
energy-momentum density, describing the corresponding Lorentzian 
theory in a thermal state. The dual of this is a Euclidean AdS
black brane, not a Lorentzian geometry.

In this context, consider a class of asymptotically de Sitter 
spacetimes
\be\label{dSbwCmplx0}
ds^2 = -{R_{dS}^2 d\tau^2\over\tau^2 (1+{C\over\tau^d})} 
+ {\tau^2\over R_{dS}^2} \Big(1+{C\over\tau^d}\Big) dw^2 
+ {\tau^2\over R_{dS}^2} dx_i^2\ , \qquad C\propto \tau_0^d\ ,
\ee
with $C$ a general complex parameter and $\tau_0$ is real. 
This metric should be regared as a (generally complex) saddle point in a functional integral. Motivated by this, we impose a requirement that the euclidean metric obtained by Wick rotation of the time coordinate 
$\tau$ is real and regular -- this fixes the parameter $C=-i^d\tau_0^d$, and
requires $w$ to be periodic. However, as will be clear in the following, the Lorentzian metric can become singular for even $d$.  Our solutions are similar to those in \cite{Anninos:2012ft} who considered solutions with $S^{d-1} \times S^1$ boundaries. In fact (\ref{dSbwCmplx0}) can be obtained as a limit of the solution \cite{Anninos:2012ft} when the radius of the $S^1$ is much smaller than the radius of $S^{d-1}$ (or equivalently, as $S^{d-1}$ decompactifies).

The resulting spacetime 
can equivalently be obtained from the Euclidean $AdS$ black brane 
by the analytic continuation from $AdS$ to $dS$ familiar in $dS/CFT$ 
\cite{Strominger:2001pn,Maldacena:2002vr}. Clearly this leads to a 
{\em complex} solution for odd $d$. This is in fact quite common in 
the dS/CFT correspondence \cite{Maldacena:2002vr,Harlow:2011ke}. Indeed 
we will show that the energy-momentum tensor 
$T_{ij}\sim {\delta\Psi\over\delta h^{ij}}$ in the CFT which follows 
from (\ref{one}) 
%\ben\label{Tij}\een
is real for odd $d$. This is consistent with known results for 
correlators in pure $dS$ \footnote{One may wonder if there could be an 
additional factor of $i$ in this relation. However, the requirement that 
the $n$-point correlator does not have a $n$-dependent phase rules this 
out \cite{Anninos:2011ui}.}. For example in $d=3$ we get
$\langle T_{ij}\rangle\propto {\tau_0^3\over G_4 R_{dS}^4}$~, %\label{six}
which is exactly what we need. For even $d$, the solution 
(\ref{dSbwCmplx0}) is real and the boundary energy-momentum tensor is 
purely imaginary.
%obtained by replacing $\tau_0^d \rightarrow i \tau_0^d$.

While real energy momentum tensors are thus obtained only for complex
solutions, it is interesting to consider the geometry of the solutions
with real parameters. The geometry is bounded by asymptotically de
Sitter spacelike $\cI^\pm$ and time-like singularities at the two ends
of space. The null lines $\tau = \tau_0$ are Cauchy horizons. In fact
the geometry bears some resemblance to the interior of the
Reissner-Nordstrom black hole \cite{mtw,hawkingEllis}. The physics of
physical observers is also quite similar. Timelike geodesics
originating from $\cI^-$ are repelled by the singularities. As an observer approaches the horizon, light coming from $\cI^-$ is
infinitely blueshifted, just as in the RN interior. It is natural to
expect that this blueshift signals an instability, preserving cosmic
censorship and distinguishing these from naked singularities. It is
intriguing to note that from a $dS/CFT$ perspective, the
energy-momentum tensor is purely imaginary. It is tempting to think of
this imaginary $T_{ij}$ as a possible dual signature of the Cauchy
horizon blue-shift instability that we have seen. It would be
interesting to explore this and more generally cosmic censorship in
$dS/CFT$. We dub these solutions ``bluewalls''.

\section{$dS/CFT$ at uniform energy-momentum density}

The CFT correlation functions in $dS/CFT$ correspondence follow
from analytic continuation from euclidean $AdS$ (or double analytic
continuation from lorentzian $AdS$), with the interpretation that the wavefunctional is the generating functional of correlators. One half of $dS_{d+1}$, e.g. the
upper patch being ${\cal I}^+$ at $\tau = \infty$ with a coordinate horizon at
$\tau = 0$ is described in the planar coordinate foliation by the metric
\ben\label{dSpoinc}
ds^2 = -R_{dS}^2{d\tau^2\over \tau^2} + \frac{\tau^2}{R_{dS}^2} 
\delta_{ij}dx^i dx^j\ .
\een
This may be obtained by analytic continuation of a Poincare 
slicing of $EAdS$, 
\be\label{AdStodS}
r \rightarrow -i\tau\ ,\qquad\ \ R_{AdS} \rightarrow -iR_{dS}\ .
\ee
In fact the analytic continuation of the smooth euclidean solutions 
lead to Bunch-Davies initial conditions on the cosmological horizon.

Consider the asymptotically de Sitter spacetime
\be\label{dSbwCmplx}
ds^2 = -{R_{dS}^2 d\tau^2\over\tau^2 (1+{C\over\tau^d})} 
+ {\tau^2\over R_{dS}^2} \Big(1+{C\over\tau^d}\Big) dw^2 
+ {\tau^2\over R_{dS}^2} dx_i^2\ ,
\ee
with $C$ a general complex parameter.
This is a {\em complex} metric which satisfies Einstein's equation with 
a positive cosmological constant 
\be\label{dSeom}
R_{MN}={d\over R_{dS}^2}g_{MN}~, \qquad \Lambda={d(d-1)\over 2R_{dS}^2}\ .
\ee
With a view to requiring an analog of regularity in the interior 
for an asymptotically $AdS$ solution, consider a Wick rotation of 
the time coordinate $\tau$ above. Then (\ref{dSbwCmplx}) becomes
\be\label{Wickrot}
\tau=il\quad \Rightarrow\quad
ds^2_E = -{R_{dS}^2 dl^2\over l^2 (1+{C\over i^dl^d})} 
- {l^2\over R_{dS}^2} \Big(1+{C\over i^dl^d}\Big) dw^2 
- {l^2\over R_{dS}^2} dx_i^2\ .
\ee
With a further continuation $R_{dS} \rightarrow iR^\prime$, this is in
general a complex euclidean metric.
We require that this euclidean spacetime is real and 
regular in the interior, by which we 
demand that the spacetime in the interior approaches flat Euclidean 
space in the $(l,w)$-plane with no conical singularity. This is true if 
\be\label{dSreg}
C = -i^d \tau_0^d\ , \qquad l\geq \tau_0\ , \qquad 
w \simeq w + {4\pi\over (d-1) \tau_0}\ ,
\ee
where $\tau_0$ is some real parameter of dimension length, and the 
$w$-coordinate is compactified with the periodicity fixed by demanding 
that there is no conical singularity.

This requirement of regularity is similar to the one we use in an
asymptotically $AdS$ spacetime, where \eg\ Wick rotating the time
coordinate renders the resulting Euclidean space regular if the time 
coordinate is regarded as compact with a periodicity that removes 
any conical singularity (thus rendering it sensible for a Euclidean 
path integral). A sharp difference in the asymptotically 
de Sitter case is that we Wick rotate the asymptotic bulk time 
coordinate but the absence of a conical singularity fixes the 
$w$-coordinate to be compact with appropriate periodicity. 
This, however, is at odds with the regularity of the real time metric
with this value of $C$ when $d$ is even. In that case, a periodic $w$
leads to a conical type singularity at $\tau = \tau_0$ pretty much
like the Milne universe with a compact spatial direction.
In this regard, it is interesting to consider the asymptotically 
$dS_5$ solution above: then the above Wick rotation procedure fixes 
$C=-\tau_0^4$ and the periodicity of the $w$-coordinate and the solution is
\be\label{dS5bwReg}
ds^2 = -{R_{dS}^2 d\tau^2\over\tau^2 (1-{\tau_0^4\over \tau^4})} 
+ {\tau^2\over R_{dS}^2} \Big(1-{\tau_0^4\over \tau^d}\Big) dw^2 
+ {\tau^2\over R_{dS}^2} dx_i^2 \ .
\ee
The metric in the vicinity of $l=\tau_0$ is\ 
%\be
$ds^2 \sim -dT^2 + T^2 dw^2 + \tau_0^2 dx_i^2$\ , where $T\sim l-\tau_0$.
This is Milne space in the $(T,w)$-plane, with $w$ compact (and thus 
a resulting singularity). We note that Wick rotating the coordinate $T$ does 
not give a Euclidean space and is not equivalent to the above procedure 
of Wick rotating the asymptotic time coordinate $\tau$.

As expected, this entire procedure is equivalent to analytically 
continuing from the Euclidean $AdS$ black brane 
\be\label{EAdSbb}
ds^2 = R_{AdS}^2\frac{dr^2}{r^2(1-\frac{r_0^{d}}{r^d})} 
+ \frac{r^2}{R_{AdS}^2}\Big( 1-\frac{r_0^{d}}{r^d}\Big) d\theta^2 
+ + \frac{r^2}{R_{AdS}^2} \sum_{i=1}^{d-1}dx^i dx^i\ ,
\ee
where $\theta \sim \theta + \frac{4\pi}{(d-1)r_0}$~, to the 
asymptotically de Sitter spacetime (\ref{dSbwCmplx}) using (\ref{AdStodS})
and we identify $r_0\equiv \tau_0$. The phase obtained by this analytic 
continuation is ${-1\over (-i)^d}$ which can be seen as identical to 
$-i^d$ in (\ref{dSreg}). The regularity criterion (\ref{dSreg}) itself 
is then seen to simply be the analog of regularity of the $EAdS$ black 
brane. The condition $l\geq \tau_0$ is 
equivalent to the radial coordinate having the range $r\geq r_0$. 
In the Lorentzian signature spacetime (\ref{dSbwCmplx}), the time 
$\tau$-coordinate extends to $\tau\ra 0$.
The curvature invariant $R_{\mu\nu\rho\sigma}R^{\mu\nu\rho\sigma}$ 
diverges as $\tau\ra 0$. 

Near $\cI^+$, \ie\ $\tau \rightarrow \infty$, the metric 
(\ref{dSbwCmplx}) approaches that of de Sitter space with a 
Fefferman-Graham expansion
\be\label{FGdS}
ds^2 = -{R_{dS}^2\over \tau^2}d\tau^2 + h_{ij} dy^i dy^j 
= -{R_{dS}^2\over \tau^2}d\tau^2 + {\tau^2\over R_{dS}^2} 
\Big[g_{ij}^{(0)}(y^i) + {R_{dS}^2\over \tau^2} 
g_{ij}^{(2)}(y^i)+\ldots\Big] dy^i dy^j\ .
\ee
It is clear from (\ref{FGdS}) that ``normalizable'' metric pieces are 
turned on in (\ref{dSbwCmplx}) \footnote{Scalar modes in $dS_{d+1}$ 
near the boundary are\ 
%$\tau^{d+1}\del_\tau(\tau^{1-d}\del_\tau\phi)+(k_i^2\tau^2+m^2)\phi=0$. 
$\phi\sim \tau^\Delta$, with\ $\Delta (\Delta-d)=-m^2R^2$. For 
$m^2=0$, we have\ $\Delta=d$ as analogous to a ``normalizable'' mode 
(in $AdS$): this is the mode turned on in (\ref{dS4whcont}).}. 
We then expect a nonzero expectation value for the energy-momentum 
tensor here, as in the $AdS$ context \cite{Balasubramanian:1999re,
Myers:1999psa,de Haro:2000xn,Skenderis:2002wp}.
For concreteness, let us consider the asymptotically $dS_4$ solution 
(\ref{dSbwCmplx}) with the regularity conditions (\ref{dSreg}),
\be\label{dS4whcont}
ds^2 = -{R_{dS}^2 d\tau^2\over\tau^2 (1+{i\tau_0^3\over\tau^3})} 
+ {\tau^2\over R_{dS}^2} \Big(1+{i\tau_0^3\over\tau^3}\Big) dw^2 
+ {\tau^2\over R_{dS}^2} dx_i^2\ .
\ee
The calculation of the energy momentum tensor proceeds in a way entirely 
analogous to that in $AdS$. The total action, obtained by adding suitable 
Gibbons-Hawking surface terms and counterterms to the bulk action is
\ben 
I = {1\over 16\pi G_4} \int_{\cal M} 
d\tau d^3x \sqrt{-g}\ (R - 2\Lambda) +\ {1\over 8\pi G_4}
\int_{\del\cal M} d^3x \sqrt{h}\ \big(K + {2\over R_{dS}}\big)\ 
\een
The counterterms have been engineered to remove divergences in the 
bulk action coming from the boundary at $\tau \rightarrow \infty$ . 
Here $h_{ij}$ is the boundary metric and $K$ is the trace of the 
extrinsic curvature. This renormalized action appears in (\ref{one}). 
This leads to the energy momentum tensor\footnote{Note that our 
definition is consistent with \cite{Maldacena:2002vr} (also 
\cite{Anninos:2011ui}), but differs from \eg\ 
\cite{Strominger:2001pn,Balasubramanian:2001nb} which use a 
derivative of the action rather than the wavefunction.}
\be\label{dS4Tij}
T_{ij} = {\displaystyle \lim_{\tau\ra\infty}}\ {\tau\over R_{dS}}\
 {2\over\sqrt{h}} {\delta \Psi\over\delta h^{ij}} 
\sim\ {\displaystyle \lim_{\tau\ra\infty}}\ {\tau\over R_{dS}} {i\over G_4}
\Big(K_{ij} - Kh_{ij} - {2\over R_{dS}} h_{ij}\Big)\ ,
\ee
We have used the standard relationship\ 
$\sqrt{h^B} h^B_{\mu\nu}T^{\nu\rho}=\sqrt{h} h_{\mu\nu}\tau^{\nu\rho}$\ 
between the energy momentum tensor of the boundary theory and the 
quasi-local stress tensor $\tau^{\mu\nu}$,
%\ben
with $h^B_{\mu\nu}=\lim_{\tau\ra\infty} {R_{dS}^2\over\tau^2} h_{\mu\nu}$ the 
boundary metric.
The above energy-momentum tensor vanishes for pure $dS_4$ as expected. 
For the spacetime (\ref{dS4whcont}), we obtain
\be\label{Tijdswh}
T_{ww} = -2T_{ii} \sim\ {i\over G_4} 
\Big({i\tau_0^3\over R_{dS}^4}\Big) = -{\tau_0^3\over G_4R_{dS}^4}\ ,
\ee
which is a real and spatially uniform energy-momentum density. Since 
(\ref{dS4whcont}) is a complex solution, its conjugate is also a 
solution (obtained by analytically continuing the opposite way), 
giving $T_{ij}$ of the opposite sign as above.
In the $AdS$ case, spacetimes of this sort which are solutions in 
pure gravity have\ $I_{bulk}={1\over 16\pi G_4} \int_{\cal M} 
drd^3x \sqrt{-g}\ (R - 2\Lambda) = {1\over 16\pi G_4} \int drd^3x 
{R^4\over r^4} ({-6\over R^2})$.\ Under the analytic continuation 
(\ref{AdStodS}), we have
\be
I_{EAdS}\ \ra\ {1\over 16\pi G_4} \int (-id\tau)d^3x {R_{dS}^4\over\tau^4}\
\Big({-6\over -R_{dS}^2}\Big) = -i I_{dS}\ ,
\ee
where $I_{dS}$ is the action for the asymptotically de Sitter 
solution\footnote{As we saw, the divergent terms in $I_{dS}$ cancel: 
this gives a single new term
%${1\over 16\pi G_4} \int_{\del\cal M} d^3x \int 
%{R_{dS}^4d\tau\over \tau^4} ({6\over R_{dS}^2})|_{\tau^3=iR_{dS}^6/r_0^3} 
%= -{iV_3r_0^3\over 8\pi G_4R_{dS}^4}$~, obtained 
from the $\tau$-location where $h_{ij}$ departs from the $dS_4$ value.
The on-shell wavefunction for these solutions becomes\ 
$\Psi \sim\ \Psi_{dS}\ exp[{V_3\tau_0^3\over 8\pi G_4R_{dS}^4}]$.}. 
Thus the energy-momentum tensor is continued as\
${2\over\sqrt{h}} {\delta (-I_{EAdS})\over\delta h^{\mu\nu}} \ra 
{2\over\sqrt{h}} {\delta (iI_{dS})\over\delta h^{ij}}$. 
Note that $T_{ij}$ in (\ref{Tijdswh}) is traceless 
($T_{ww}+2T_{ii}=0$)\ as expected for a CFT.

Thus this asymptotically $dS_4$ complex solution is dual to a
euclidean CFT with spatially uniform energy-momentum density, \ie\
uniform $T_{ij}$ expectation value (\ref{Tijdswh}) in the Euclidean
CFT dual to $dS_4$ (in Poincare slicing). Loosely speaking, this
Euclidean partition function corresponds to a thermal state of the 
would-be corresponding Lorentzian theory (on $R\times R^2$), analogous 
to the $AdS_4$ Schwarzschild black brane dual to a thermal state in 
the SYM CFT with uniform $T_{\mu\nu}$.

Similar arguments apply in other dimensions but with different results. 
Using the Fefferman-Graham expansion (\ref{FGdS}) for an asymptotically 
$dS_{d+1}$ spacetime, we see that ``normalizable'' metric modes 
$g^{(d)}_{\mu\nu}$ turned on give rise to a nonzero expectation value 
for the holographic energy-momentum tensor 
\be\label{dSd+1Tij}
T_{ij} = {\displaystyle \lim_{\tau\ra\infty}}\ 
{\tau^{d-2}\over R_{dS}^{d-2}}\
 {2\over\sqrt{h}} {\delta \Psi\over\delta h^{ij}} 
\sim\ {\displaystyle \lim_{\tau\ra\infty}}\ {\tau^{d-2}\over R_{dS}^{d-2}} 
{i\over G_{d+1}} \Big(K_{ij} - Kh_{ij} - {d-1\over R_{dS}} h_{ij}\Big) 
\propto {i\over G_{d+1} R_{dS}} g^{(d)}_{ij}\ ,
\ee
where the form of (\ref{FGdS}) shows $g^{(d)}_{ij}$ to be the 
dimensionless coefficient of the normalizable ${1\over\tau^{d-2}}$ term, 
and the $i$ arises from $\Psi$, the wavefunction of the universe.
In effect, this $dS/CFT$ energy-momentum tensor can be thought of as 
the analytic continuation of the $EAdS$ one, with the $i$ 
arising from $R_{AdS}\ra -iR_{dS}$, and the metric modes also continuing
correspondingly. The spacetime (\ref{dSbwCmplx}) with the parameter 
$C=-i^d\tau_0^d$ in (\ref{dSreg}) gives
\be\label{dSd+1Tij2}
g^{(d)}_{ww} \sim -{i^d\tau_0^d\over R_{dS}^d} \qquad\Rightarrow\qquad
T_{ww} \sim\ -{i^{d+1}\tau_0^d\over G_{d+1} R_{dS}^{d+1}}\ =\ 
{i^{d-1}\tau_0^d\over G_{d+1} R_{dS}^{d+1}}\ ,
\ee
with $T_{ww}+(d-1)T_{ii}=0$. The phase $i^{d-1}$ is equivalent to that in 
general $dS/CFT$ correlation functions arising from the analytic continuation 
(\ref{AdStodS}) from $EAdS$ correlators \cite{Das:2013qea}, following the arguments of \cite{Anninos:2011ui}. For even $d$ the energy momentum tensor is imaginary -- this is also the case when the Lorentzian signature metric is singular at $\tau = \tau_0$.

We thus see that a real energy-momentum density must arise from a 
metric mode $g^{(d)}$ that is pure imaginary: in other words, the 
spacetime (\ref{dSbwCmplx}) with a pure imaginary parameter $C$ is dual 
to a CFT with real spatially uniform energy-momentum density. 
For the $dS_5$ case, we see that the regularity criterion for the Euclidean solution (or 
equivalently the analytic continuation from the $EAdS$ black brane) 
gives the spacetime (\ref{dS5bwReg}) which is real: this metric which is singular gives an 
imaginary $T_{ij}$ above. 
%A real density $T_{ij}$ is obtained by 
%considering a spacetime (\ref{dSbwCmplx}) with pure imaginary $C$.

In summary, we have described asymptotically de Sitter spacetimes (\ref{dSbwCmplx}) 
which under a Wick rotation are regular in the interior for certain 
values of the general complex parameter (\ref{dSreg}). The resulting 
spacetime can then be equivalently obtained by analytic continuation 
(\ref{AdStodS}) from the Euclidean $AdS$ black brane (\ref{EAdSbb}).
These spacetimes give rise to a spatially uniform holographic 
energy-momentum density (\ref{dSd+1Tij}), which is real if the 
spacetime is complex (for odd $d$). 
Conversely, given a $T_{ij}$ expectation value in $dS/CFT$, we could ask 
what the gravity dual is. An asymptotically de Sitter spacetime with the 
Fefferman-Graham series expansion (\ref{FGdS}) and thus corresponding 
$T_{ij}$ then in fact sums to the closed form expression (\ref{dSbwCmplx}).

\section{Real parameter $C$:\ \ a de Sitter ``bluewall''}

Even though the metric needs to be complex to yield a real energy momentum 
tensor, it is interesting to explore the properties of metrics of the form 
(\ref{dSbwCmplx}), (\ref{dSreg}), but with the parameter $\tau_0^d$ also 
continued to be real, \ie\
\be\label{dSwh}
ds^2 = -{d\tau^2\over f(\tau)} 
+ f(\tau) dw^2 + \tau^2 dx_i^2\ ,
\qquad f(\tau)= \tau^2 \Big(1-{\tau_0^d\over\tau^d}\Big)\ ,
\ee
with a nonzero parameter $\tau_0$, and $x_i$ are $d-1$ of the $d$ 
spatial dimensions. The $w$-coordinate here has the range 
$-\infty\leq w\leq \infty$. This can be recast in FRW form as an 
asymptotically deSitter cosmology with anisotropy in the $w$-direction. 
The metric (\ref{dSwh}) is simply the analytic continuation of AdS-Schwarzschild with a further continuation of the mass parameter. We do not speculate about the significance of this real solution for $dS/CFT$ for odd $d$.

\begin{figure}[h] 
\hspace{1pc} \includegraphics[width=11pc]{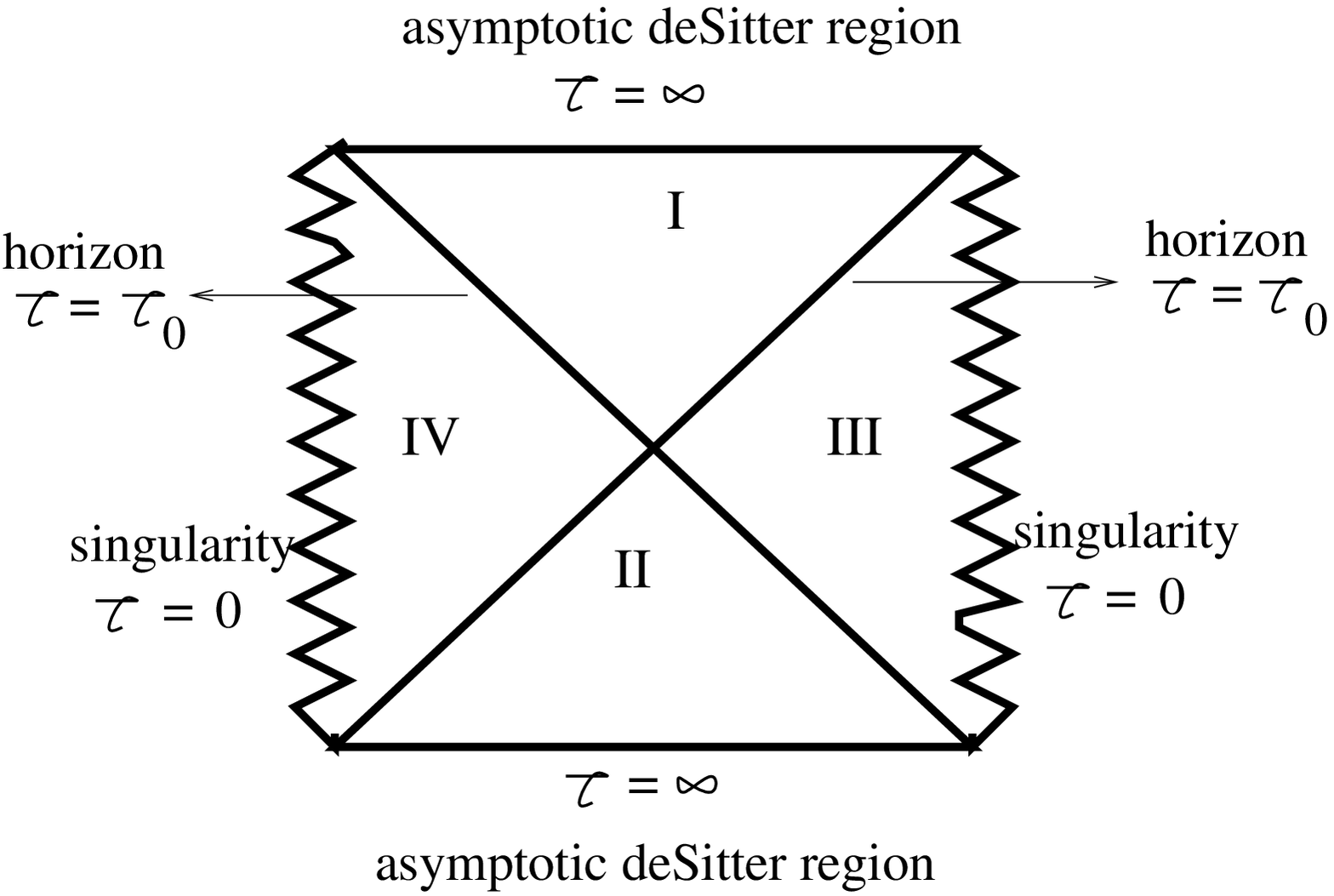} \hspace{2pc}
\begin{minipage}[b]{20pc}
\caption{{\label{figdS}\footnotesize {de Sitter ``bluewall'' Penrose diagram.
\newline 
This resembles the Penrose diagram of the \newline 
$AdS$ Schwarzschild black brane rotated by ${\pi\over 2}$. \newline \newline }}}
\end{minipage}
\end{figure}

The lines $\tau = \tau_0$ are coordinate singularities whose nature will 
be explored below. For concreteness, we focus on $d=3$. The maximally 
extended geometry in Kruskal type coordinates (Appendix A, 
eq.(\ref{Kruskaluv})) is 
\be\label{metKruskal}
ds^2 = \tau^2 \left[ -{4\over 9} 
\Big(1+{\tau_0\over\tau}+{\tau_0^2\over\tau^2}\Big)^{3/2} 
e^{-\sqrt{3}\tan^{-1}({2{\tau\over\tau_0}+1\over \sqrt{3}})} 
d{\tilde u} d{\tilde v} + dx_i^2 \right] .
\ee
The Penrose diagram\footnote{The Penrose diagram Figure~\ref{figdS} 
also appears in \cite{Astefanesei:2003gw} but corresponds to a distinct 
spacetime (with an inhomogenuous energy-momentum tensor).} Figure~\ref{figdS} shows the following key features 
of the geometry.

\noindent {\bf \emph{Two asymptotic $dS$-regions:}}\ \
$v^2-u^2={\tilde u}{\tilde v}>0$ both map to $\tau\gg \tau_0$, 
using (\ref{Kruskaluv}).

\noindent {\bf \emph{Cauchy\ horizons:}}\qquad
$\tau=\tau_0\quad \Rightarrow\quad {\tilde u} {\tilde v} = 0\ ,\ \ \ie\ \ \
u=\pm v$.\\
%There are two horizons, at ${\tilde u}=0$ and ${\tilde v} = 0$. 
Using (\ref{Kruskaluv}), we see that\ $\tanh {3w\tau_0\over 2} = 
{{\tilde u}-{\tilde v}\over {\tilde u}+{\tilde v}}$\ so that the two 
horizons are\ 
${\tilde u} = 0  \Rightarrow  \tau=\tau_0,\ w=-\infty$, and\
${\tilde v} = 0  \Rightarrow \tau=\tau_0,\ w=+\infty$.
These are Cauchy horizons, as we discuss later.\ 
We refer to the intersection of the horizons\ 
${\tilde u} = 0 = {\tilde v}$, \ie\ 
$u=0=v$\ or $\tau=\tau_0$\ as the {\bf bifurcation region}: the 
$w$-coordinate can take any value here.

\noindent {\bf \emph{Timelike\ singularities:}}\qquad
%The lines \ben
$\tau=0\quad \Rightarrow\quad 
{\tilde u} {\tilde v} = -e^{\pi\over 2\sqrt{3}}\ \sim\ v^2 - u^2$\\
In the Kruskal diagram these are hyperbolae with 
$u^2 > v^2$. There are two singularity loci\ 
${\tilde v} = -{c\over {\tilde u}}$\ with ${\tilde u}>0$ and ${\tilde u}<0$.\
The curvature invariants for (\ref{dSwh}) are 
\bc
$R=d(d+1)\ ,\quad R_{\mu\nu}R^{\mu\nu}=d^2(d+1)\ ,\quad 
R_{\mu\nu\rho\sigma}R^{\mu\nu\rho\sigma} 
= 2d \Big(d+1 + {(d-2) (d-1)^2\over 2}  {\tau_0^{2d}\over\tau^{2d}}\Big).$
\ec
The divergence in $R_{\mu\nu\rho\sigma}R^{\mu\nu\rho\sigma}$ implies a 
curvature singularity as $\tau\ra 0$: this is what the Schwarzschild interior singularity becomes after analytic continuation.
Interestingly, the $dS_3$ solution 
 is singularity-free: the metric in this case is isomorphic to 
$dS_3$ in static coordinates. 

Near $\tau\ra 0$, the metric approaches\ 
$ds^2 \sim\ -{dw^2\over \tau^{d-2}} + \tau^{d-2} d\tau^2 + {dx_i^2\over\tau^2}
\ \sim\ {1\over\tau} d{\tilde u} d{\tilde v} + \tau^2 dx_i^2$.
For $\tau<\tau_0$, the $\tau$-coordinate is spacelike while $w$ is 
timelike. Then the singularity which occurs on a constant-$\tau$ slice 
is timelike (metric approaching $ds^2\sim -{dw^2\over \tau^{d-2}}$).

Note that these features (Cauchy horizons, timelike 
singularities) resemble the interior of the Reissner-Nordstrom black 
hole or ``wormhole'' (discussed in \eg\ \cite{hawkingEllis}). Recall 
that the latter geometry is of the form
\ben
ds^2 = -f(r) dt^2 + {dr^2\over f(r)} + r^2d\Omega^2\ ,\qquad
f(r)=(r-r_+)(r-r_-)\ .
\een
Near the inner horizon $r_-$ this can be approximated as\
$ds^2\sim\ -{dr^2\over k(r-r_-)} + k(r-r_-) dt^2 + r^2d\Omega^2$,
where $k=r_+-r_-$. In the region $r_-<r<r_+$, the radial 
coordinate $r$ is timelike. Thus we see that the geometry near the 
inner horizon $r_-$ in fact resembles the geometry\ 
$ds^2 \sim\ -{d\tau^2\over\tau-\tau_0} + (\tau-\tau_0)dw^2+\tau^2dx_i^2$\
near the horizon $\tau_0$ in the present $dS$-case. Thus it is not 
surprising that the Penrose diagram and associated physics are 
similar in both cases.

For general timelike geodesic trajectories the momenta satisfy $p_\mu p^\mu = -m^2$ and the action is
\ $S = \int d\tau {m\over 2} g_{\mu\nu} {\dot x}^\mu {\dot x}^\nu$\ ,
with $\lambda$ the affine parameter. Sinc $\partial_w$ is  a Killing vector the associated momentum $p_w$ is conserved
\be\label{geodTimelike}
{p_w\over m}=\Big(\tau^2-{\tau_0^3\over\tau}\Big) {dw\over d\lambda}\ , \quad
{{\dot\tau}^2 -\ p_w^2/m^2 \over \tau^2 -{\tau_0^3\over\tau}}
%\over\tau^2 -{\tau_0^3\over\tau}}
\ = 1 , \quad\ %\nonumber\\
{dw\over d\tau} = {\pm {p_w\over m}\over\tau^2 
(1-{\tau_0^3\over\tau^3}) \sqrt{{p_w^2\over m^2} + 
\tau^2 (1-{\tau_0^3\over\tau^3})}}\ .
\ee
In these coordinates static observers are timelike geodesics at 
const-$w,x_i$ with\ 
$p_w=0 ,\ g_{\tau\tau} ({d\tau\over d\lambda})^2 = -1$. Using 
(\ref{Kruskaluv}), these are\ ${u\over v}=-\tanh {3w\tau_0\over 2}=const$ 
\ie\ straight lines crossing from the past universe $II$ to the future 
one $I$ through the bifurcation region.
Generic observers have $p_w\neq 0$: as $\tau\ra\tau_0$, they approach 
the horizon with increasing coordinate speed\ 
$|{dw\over d\tau}|\ra\infty$ and fall through the horizon. 
They do not hit the singularity however: the 
singularity appears to be repulsive. This can be seen from 
(\ref{geodTimelike}) by noting that\ 
${\dot\tau}^2 = {p_w^2\over m^2} + \tau^2 - {\tau_0^3\over\tau} > 0$\ 
implies a turning point $\tau_{min}={\tau_0^3\over p_w^2}$, and likewise\
$({dw\over d\tau})^2>0\ \Rightarrow\ \tau\geq\tau_{min}$,
so that timelike geodesic trajectories never reach the singularity.
We see that in this ``deSitter bluewall'' solution, particles 
can apparently pass from the past universe through the horizons 
avoiding the timelike singularities behind the horizons and emerge 
in the future universe. Whether such trajectories can actually 
go across is unclear due to a blue-shift instability stemming from 
the Cauchy horizons, as we discuss now.

\noindent {\bf \emph{Bluewalls:}}\ \ 
We now discuss the role of the Cauchy horizons and the possibility
of traversing from the past universe to the future one.
%The discussion below is quite similar to that for the Reissner-Nordstrom 
%black hole near the inner horizon $r_-$.\
First recall that the horizons are ${\tilde u}=0$ at $\tau=\tau_0,
w=-\infty$ and ${\tilde v}=0$ at $\tau=\tau_0, w=+\infty$. Thus the
horizons are actually infinitely far away in the $w$-direction. As 
we have seen, trajectories from the past universe (beginning at some
point on $\cI^-$) can pass through the horizons into the interior
regions: however there are timelike and null geodesics which begin 
in the interior
regions alone and thus cannot be obtained by time development of any
Cauchy data on $\cI^-$. Thus the past horizons are future Cauchy 
horizons for Cauchy data on $\cI^-$. Likewise the future horizons 
are causal boundaries for the future universe, so that these are 
past Cauchy horizons for data on $\cI^+$.

\begin{figure}[h] 
\hspace{0.5pc} \includegraphics[width=13pc]{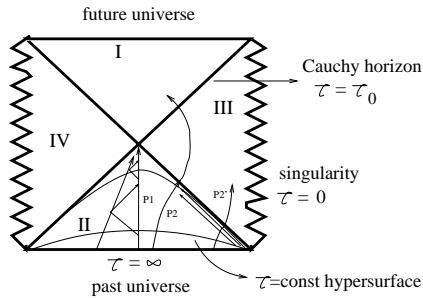} \hspace{1pc}
\begin{minipage}[b]{24pc}
\caption{{\label{figdSpaths}\footnotesize {Trajectories in the de Sitter 
bluewall and the Cauchy horizon. Observers $P_1$ are static while $P_2$ 
has $w$-momentum $p_w$, crosses the horizon, turns around inside and 
appears to re-emerge in the future universe. 
%Observers $P_1$ exchange light signals between them as they pass through the bifurcation region. 
Also shown are incoming lightrays from infinity which ``crowd near'' 
the Cauchy horizon.\newline }}}
\end{minipage}
\end{figure}
Two static observers at different $w$-locations communicating 
by lightray signals are always in contact with each other.
Consider observers $P_1, P_2$, with $P_1$ a static observer while
$P_2$ is a geodesic infalling observer with some $w$-momentum. $P_2$
falls freely through the horizon, turns around somewhere in the
interior and then appears to re-emerge in the future universe. From
the point of view of $P_1$, the observer $P_2$ appears to be going to
$|w|\ra\infty$.
%$P_2$ can receive a signal from $P_1$ but the time
%lapse between signals becomes successively longer. 
Eventually $P_2$ sends, from $|w|\ra\infty$, a ``final'' lightray which is the
generator of the corresponding horizon. Similarly one can consider
signals received by infalling observers $P_2$ at late times, sent by
infalling observers $P_{2'}$ at early times. Such observers $P_2, P_{2'}$ 
have\ ${\dot\tau}^2 \equiv ({d\tau\over d\lambda})^2 
= {p_w^2\over m^2} + \tau^2 - {\tau_0^3\over\tau}$,\ using 
(\ref{geodTimelike}), so that we have the proper time intervals 
\be
\Delta\lambda_{P_2}\sim {1\over p_w} \Delta\tau\qquad ({\rm near}\ \tau_0)\ ,
\qquad\quad
\Delta\lambda_{P_{2'}}\sim {\Delta\tau\over\tau}\qquad ({\rm early\ times})\ .
\een
Ingoing lightray congruences of the form\ $U=w-\tau_*=c$ have a 
cross-sectional vector $v=w+\tau_*$. 
%and a cross-sectional width 
%$\Delta v = 2\Delta\tau_* = {2\Delta\tau\over \tau-\tau_0}$.
%Thus $\tau-\tau_0\sim e^v$ for such lightrays coming in from infinity.
%The  Thus\ 
%$\Delta\lambda \sim\ \Delta\tau \sim e^v \Delta v\ra 0$ as\ $\tau\ra\tau_0$,
%so that a fixed set of lightrays (fixed $\Delta v$) reaches the 
%infalling observer in a vanishingly small proper time interval.
To analyse these transmitting-receiving events further, it is convenient 
to use Eddington-Finkelstein-type coordinates here: defining the ingoing 
coordinate\ $v=w+\tau_*$, the metric (\ref{dSwh}) becomes
\be
ds^2 = f(\tau) dv^2 - 2dvd\tau + \tau^2dx_i^2\ ,
\ee
and infalling geodesic observers at const-$x_i$ (\ie\ $\tau, w$ 
decreasing with proper time $\lambda$) are 
\be
f{\dot v}^2 - 2{\dot v}{\dot \tau} = -1\ ,\qquad  
-f{\dot v}+{\dot \tau} = -f{\dot w} = p_v>0  \qquad\ (\ie\ {\dot w}<0)\ ,
\ee
\be
\Rightarrow\ \  {d\tau\over d\lambda} = -\sqrt{f(\tau)+p_v^2}\ < 0\ ,
\ \ \ {d v \over d\lambda} = - {p_v+\sqrt{f(\tau)+p_v^2}\over f(\tau)}\ ,\quad
{dv\over d\tau} = {p_v+\sqrt{f(\tau)+p_v^2}\over f(\tau) 
\sqrt{f(\tau)+p_v^2}}\ .
\label{26}
\ee
Figure~\ref{figdSpaths} shows infalling observers $P_2$ approaching 
the horizon, receiving at late times ($\tau\sim\tau_0$) light signals 
that emanate from early times ($\tau\sim\infty$): the latter can be 
thought of as signals transmitted by infalling observers $P_{2'}$ at 
early times. It can be seen from Figure~\ref{figdSpaths} that such 
events (transmission-reception of such light signals) are consistent 
with the causal (lightcone) structure of the spacetime. The light rays
in question have constant $v$ which is very large and negative.

Let us denote the conserved momenta for the two geodesics $P_2$ and
$P_{2'}$ by $p_v$ and $p_v'$ respectively.  Suppose $P_{2'}$ sends out
successive light signals along constant $v$ and constant $v + dv$, at
coordinate times $\tau^\prime$ and $\tau^\prime + d\tau^\prime$, and
these are received by $P_2$ at coordinate times  $\tau$ and 
$\tau + d\tau$. The
proper time between emission these signals is $d\lambda_{P_2'}$ while
the proper time between reception of the same two signals
by $P_2$ is $d\lambda_{P_2}$. Then equations (\ref{26}) yield
\ben
\frac{d\lambda_{P_2}}{d\lambda_{P_2'}} = \frac{f(\tau)}{f(\tau^\prime)}
\frac{p_v^\prime + \sqrt{f(\tau^\prime) + (p_v^\prime)^2}}
{p_v + \sqrt{f(\tau) + p_v^2}}
\label{bs1}
\een
When both observers are at rest, $p_v = p_v^\prime = 0$ this leads to
the standard formula for the gravitaional redshift/blueshift. In our
setup $\tau \sim \tau_0$ while $\tau^\prime \rightarrow \infty$, so
that
\ben
f(\tau) \sim 3\tau_0 (\tau - \tau_0)\ , \qquad f(\tau^\prime) \sim
(\tau^\prime)^2\ ,
\label{bs2}
\een
which leads to 
\ben
\frac{d\lambda_{P_2}}{d\lambda_{P_2'}} \sim\ 
\frac{3\tau_0(\tau - \tau_0)}{2p_v\tau^\prime}\ .
\label{bs3}
\een
We now need to express the ratio in (\ref{bs3}) in terms of $v$. It is
clear from the last equation in (\ref{26}) that at early times the
geodesic $P_{2'}$ is described by
\ben
v(\tau^\prime) = -\frac{1}{\tau^\prime} + c^\prime\ ,
\label{bs4}
\een
where $c^\prime$ is a constant. Note that we are considering light
rays which have $v \sim -\infty$, so that the integration constant 
$c^\prime$ must be large and negative (since 
$(v-c^\prime)=-{1\over\tau^\prime}$ is small as $\tau^\prime\ra\infty$). 
The trajectory $P_2$ is described in the vicinity of $\tau = \tau_0$ by 
\ben 
\tau - \tau_0 \sim A \exp \Big[\frac{3\tau_0}{2}v\Big]\ ,
\label{bs5}
\een
where $A$ is a finite constant of integration. Substituting
(\ref{bs4}) and (\ref{bs5}) in (\ref{bs3}) we get
\ben
\frac{d\lambda_{P_2}}{d\lambda_{P_2'}} \sim\ 
-\frac{3A\tau_0}{2p_v}\ (v-c^\prime) \exp\Big[\frac{3\tau_0}{2}v\Big]\ .
\label{bs6}
\een
Thus for a fixed proper time between the signals during emission, the
proper time interval for reception becomes {\em exponentially small}
as $v \rightarrow -\infty$.

It is interesting to compare the situation with pure de Sitter. Here
$f(\tau) = \tau^2$ for all $\tau$ and the cosmological horizon is at
$\tau = 0$. In this case the ratio of the proper
time interval (\ref{bs1}) for the observers $P_2$ and $P_{2'}$ 
becomes, instead of (\ref{bs3}),
\ben
\frac{d\lambda_{P_2}}{d\lambda_{P_2'}} \sim\ 
\frac{\tau^2}{2p_v\tau^\prime}\ .
\label{bs7}
\een
Near $\tau=0$, the trajectory $P_2$ can be obtained by solving the third
equation in (\ref{26}) with $f(\tau) = \tau^2 \sim 0$. Since $U$ is
finite it is easy to see that one needs $p_v \neq 0$ and one gets
\ben
v(\tau) = -\frac{2}{\tau} + a\ ,
\label{bs8}
\een
where the constant of integration $a$ is finite. The trajectory
$P_{2'}$ is exactly the same as (\ref{bs4}).  Using this, the equation
(\ref{bs7}) becomes
\ben
\frac{d\lambda_{P_2}}{d\lambda_{P_2'}} \sim\
-\frac{2(v-c')}{(v-a)^2p_v} \sim\ -\frac{2(v-c')}{v^2p_v}\ ,
\label{bs9}
\een
where in the second equation above we have used finiteness of $a$.
Once again\ $\frac{d\lambda_{P_2}}{d\lambda_{P_2'}} \rightarrow 0$, 
however in a power law fashion. This is a much milder blueshift 
than what is experienced for our bluewall solution.

This exponentially vanishing blueshift is a reflection of the 
``crowding'' of lightrays near the horizon. 
The energy flux that the infalling observer measures 
is\ $T_{\mu\nu}v^\mu v^\nu \sim T_{vv} {\dot v}^2$. From above, we see 
that the infalling observer thus crosses a diverging flux of incoming 
lightrays in finite proper time as he approaches the 
horizon\footnote{It is a reasonable assumption that $T_{vv}$ for the 
lightrays follows a power law in $v$, akin to \cite{rp:ns}. From 
(\ref{26}), (\ref{bs5}), we have 
$\dot{v}\sim e^{-(3\tau_0/2) v}|_{v\ra -\infty} \to \infty$. Thus 
$T_{vv}\dot{v}^2$ diverges as $\tau \to \tau_0$.}, suggesting 
an instability. This is somewhat akin to the Reissner-Nordstrom black 
hole inner horizon (see \eg\ \cite{Burko:1997nz}) where an infalling 
observer receives signals from the exterior region in vanishingly small 
proper time (``seeing entire histories in a flash''). However note 
that here, this occurs for the late time infalling 
observer only as he approaches the horizon and only 
from signals emanating at early times from ``infinity'' ($|w|\ra\infty$).
Now applying the energy-momentum calculation earlier gives an
imaginary energy density $\langle T_{ij}\rangle$: it is interesting 
to ask if this is the dual CFT signature of the blue-shift 
instability. It would be interesting to explore these further, 
perhaps keeping in mind black holes, firewalls and entanglement 
\cite{Mathur:2009hf,Almheiri:2012rt,Maldacena:2013xja}.

\vspace{5mm}

{\footnotesize \noindent {\bf Acknowledgements:}\ \
It is a pleasure to thank T. Hartman, J. Maldacena, G. Mandal and
S. Trivedi for useful discussions.  KN thanks the hospitality of the
Physics Dept U. Kentucky, the IAS Princeton and the
organizers of the ``Great Lakes Strings'' conference, U. Kentucky,
USA, and ``The Information Paradox, Entanglement and Black holes''
workshop, ICTS, Bangalore, during the course of this work. SRD and KN
thank the Dept of Theoretical Physics, TIFR, for hospitality at the
incipient and final stages of this work. The work of DD and SRD is
partially supported by the National Science Foundation grant
NSF-PHY-1214341. The work of KN is partially supported by a Ramanujan
Fellowship, DST, Govt of India.}

%\vspace{-3mm}

\appendix
\section{de Sitter ``bluewall'' details}

%\be\label{dS4wh}
We give some details on the $dS_4$-''bluewall''\
${ds^2\over R^2} = -{d\tau^2\over\tau^2 (1-{\tau_0^3/\tau^3})} 
+ \tau^2 (1-\tau_0^3/\tau^3) dw^2 + \tau^2 dx_i^2$\ here.
%\ee
We can analyse the vicinity of $\tau=\tau_0$ as for the Schwarzschild 
black hole, defining a ``tortoise'' $\tau$-coordinate:
for the $dS_4$-solution, this is
\be\label{tau*}
\tau_* = \int {d\tau\over \tau^2 (1-{\tau_0^3\over\tau^3})} 
= {1\over 3\tau_0} \Big( \log {\tau-\tau_0\over
\sqrt{\tau^2+\tau\tau_0+\tau_0^2}}\ 
+\ \sqrt{3} \tan^{-1} {2{\tau\over\tau_0}+1\over\sqrt{3}}\Big) .
\ee
Analogs of Kruskal-Szekeres coordinates can then be defined as
\be\label{Kruskaluv}
{\tilde u} = e^{3(\tau_*-w)\tau_0/2}\ ,\quad 
{\tilde v} = e^{3(\tau_*+w)\tau_0/2}\ ,\qquad
{\tilde u} {\tilde v} = e^{3\tau_* \tau_0} = {\tau-\tau_0\over
\sqrt{\tau^2+\tau\tau_0+\tau_0^2}}\
e^{\sqrt{3}\tan^{-1} {2{\tau\over\tau_0}+1\over \sqrt{3}}}\ ,
\ee
and\  %\be\label{uvcoords}
$u={\tilde u}-{\tilde v}=-2e^{3\tau_*\tau_0/2} \sinh {3w\tau_0\over 2} ,\ \
v={\tilde u}+{\tilde v}=2e^{3\tau_*\tau_0/2} \cosh {3w\tau_0\over 2}$~,
giving (\ref{metKruskal}) and the Penrose diagram Figure~\ref{figdS}.
With\ $T = \int d\tau/(\tau\sqrt{1-\tau_0^3/\tau^3})$~, this 
is recast in FRW-form as an accelerating cosmology\ 
$ds^2 = -dT^2 + e^{-2T} (e^{3T}+\tau_0^3)^{4/3} 
{(e^{3T}-\tau_0^3)^2\over (e^{3T}+\tau_0^3)^2} dw^2 
+ e^{-2T} (e^{3T}+\tau_0^3)^{4/3} dx_i^2$\ with $w$-anisotropy. Further 
redefining $T=\log \eta$, we can obtain 
a Fefferman-Graham expansion for this asymptotically-$dS_4$ spacetime 
near the boundary $\tau\ra\infty$.
%$ds^2 = -{d\eta^2\over \eta^2} + (\eta^2 - {8\tau_0^3\over 3\eta} + 
%O(\eta^{-4})) dw^2 + (\eta^2 + {4\tau_0^3\over 3\eta} + O(\eta^{-4})) dx_i^2$
%Consider (\ref{dS4wh}) \ie\ the solution (\ref{dSwh}) for the 
%$dS_4$-deformation, with $\tau\ra\infty$ the asymptotic $dS_4$ boundary.
%There are in fact two asymptotic de Sitter regions here: to see this 
Following Einstein-Rosen's description \cite{Einstein:1935tc} of 
the ``bridge'' in the Schwarzschild black hole (using $\rho^2=r-2m$), 
define\ $t^2=\tau-\tau_0$. This coordinate has the range 
$t:-\infty\ra \infty$ as $\tau:\infty\ra\tau_0$ and then 
$\tau:\tau_0\ra\infty$, giving two $t$-sheets of the asymptotic deSitter 
region,
\be
ds^2 = {-4(t^2+\tau_0) dt^2 \over (t^2+\tau_0)^2+\tau_0(t^2+\tau_0)+\tau_0^2} 
+ \big({(t^2+\tau_0)^2+\tau_0(t^2+\tau_0)+\tau_0^2\over t^2+\tau_0}\big) 
t^2 dw^2 + (t^2+\tau_0)^2 dx_i^2. \nonumber
\ee
Thus the two asymptotic universes are connected by a timelike 
Einstein-Rosen bridge.
At $\tau=\tau_0$, we have $g_{ww}=0$ so the $w$-direction shrinks 
to vanishing size. Near $\tau=\tau_0$, the metric is approximated as\
$ds^2 \sim\ -{d\tau^2\over k(\tau-\tau_0)} + (\tau-\tau_0) \tau_0^2 dw^2 
+ \tau_0^2 dx_i^2\sim\ -dt^2 + t^2 d{\tilde w}^2 + d{\tilde x_i}^2$,\ 
which is flat space with the $(t,w)$-plane in Milne coordinates.

Null geodesics $ds^2=0$, in the $(\tau,w)$-plane defining lightcones and 
causal structure, are\
%\be\label{nullrays}
$dw = \pm d\tau_* ,\ \
{dw\over d\tau} = \pm {1\over \tau^2 (1-\tau_0^3/\tau^3)}$~,
with $\tau_*$ given in (\ref{tau*}). Near the horizon, the trajectories 
approach $w=\pm \tau_*+const \sim \pm\ {1\over 3\tau_0} 
\log {|\tau-\tau_0|\over 3\tau_0}$, \ie\ $w\ra \pm\infty$.
These null rays intersect the horizon and hit the singularity in the 
interior at $w_0\pm {1\over 3\tau_0} \log 3$.\
%Lightcones are defined by these null rays: generic timelike 
%paths lie in the interior of the lightcones, satisfying\ 
%$\left|{dw\over d\tau}\right| < {1\over \tau^2 (1-{\tau_0^3\over\tau^3})}$.
%Near the horizon $\tau\ra\tau_0$, the lightcones open up indefinitely.
%(contrast with the Schw.BH. where $|{dr\over dt}| < 1-{2m\over r}\ra 0$ 
%means lightcones close up near the horizon).

Note that $\tau=const$ surfaces are spacelike hyperbolic hypersurfaces 
with $v^2-u^2=const$ in the region outside the horizons, using 
(\ref{Kruskaluv}). In these exterior regions, 
\be\label{constwpath}
w=const\ \ {\rm path}\quad \Rightarrow\quad 
{u\over v}=-\tanh {3w\tau_0\over 2}=const\ ,\quad \ie\ \ \
{{\tilde u}\over {\tilde v}} 
= {1-\tanh {3w\tau_0\over 2}\over 1+\tanh {3w\tau_0\over 2}} \equiv k\ ,
\ee
\ie\ straight lines passing through the bifurcation region, 
crossing over from the past asymptotic region $II$ to the future one 
$I$. The induced worldline metric on such a $w,x_i=const$ trajectory 
and associated proper time are\
%\be\label{propT}
$dl^2 = {d\tau^2\over \tau^2-\tau_0^3/\tau}\ \equiv dT^2 ,\ \
T = {2\over 3} \log (\tau^{3/2} + \sqrt{\tau^3-\tau_0^3})$.\
The spatial metric on a $\tau=const$ hypersurface orthogonal to 
these constant-$w,x_i$ trajectories is\
${d\sigma^2\over R^2} = %-{d\tau^2\over\tau^2 (1-{\tau_0^3\over\tau^3})} 
\tau^2 \left(1-{\tau_0^3\over\tau^3}\right) dw^2 + \tau^2 dx_i^2$.\ 
We see that at the bridge $\tau\ra\tau_0$, the spatial metric 
degenerates and the cross-sectional 3-area\ 
$V_{w,x_1,x_2}=\Delta w\Delta^2x_i \tau^3\sqrt{1-\tau_0^3/\tau^3~}$ 
vanishes. 
The proper time $T$ is consistent with the equations for timelike
geodesics at constant $x_i$ in the $(\tau,w)$-plane, and is finite 
along such geodesic paths between the horizon
$\tau=\tau_0$ and any point $\tau<\infty$. It can be seen by studying 
geodesic deviation for a congruence of such timelike geodesic static 
observers with const-$w,x_i$ that there are no diverging tidal forces 
as one crosses the bifurcation region from the past universe to the 
future one.

Consider now the spacetime in Kruskal form (\ref{metKruskal}) written 
as\ $ds^2 = -2f({\tilde u},{\tilde v}) d{\tilde u} d{\tilde v} + 
g({\tilde u},{\tilde v}) dx_i^2$. A family of generic timelike paths 
in the Penrose diagram is\ ${\tilde u} = k{\tilde v} + c$,\
%\be\label{traj}
%\qquad \equiv\qquad u = {k-1\over k+1} v + {2c\over k+1} = k' v + c'\ ,
which are obtained by translating sideways the $w=const$ paths 
(\ref{constwpath}). For $c=0$, these are geodesics passing through 
the bifurcation region (without intersecting the horizon). 
Parametrizing these timelike paths as\ $x^\al(\lambda)$ in the 
${\tilde u},{\tilde v}$-plane ($x_i=const$), it can be shown that 
the acceleration components\ $a^{{\tilde u}}={\ddot{\tilde u}} +
\Gamma^{{\tilde u}}_{\al\beta} {\dot x^\al}{\dot x^\beta}$ and similarly 
$a^{{\tilde v}}$ are finite as $\tau\ra\tau_0$, as is the covariant 
acceleration norm $g_{\mu\nu}a^\mu a^\nu$. Any arbitrary smooth 
timelike trajectory can be approximated as a straight line in the 
neighbourhood of any point, in particular near the horizon. Thus the 
acceleration vanishes for any timelike path crossing the horizons. 
This is perhaps not surprising since the near horizon geometry is 
essentially Milne.

\end{document}